\documentclass[%
 aip,
 amsmath,amssymb,
 reprint,
 citeautoscript
]{revtex4-1}

\usepackage{graphicx}
\usepackage{amsmath,amsfonts,amssymb}
\usepackage{color}
\usepackage{subfigure}
\usepackage{physics}
\usepackage{dsfont}
\usepackage{hyperref}
\usepackage{url}
\usepackage{leftidx}
\usepackage{textcomp}
\usepackage{gensymb}
\usepackage{array}

\newcolumntype{M}[1]{>{\centering\arraybackslash}m{#1}}

\usepackage[utf8]{inputenc}
\usepackage[dvipsnames]{xcolor}
\usepackage{multirow}
\usepackage{comment}
\usepackage{pifont}% http://ctan.org/pkg/pifont

\usepackage{colortbl}

\usepackage{xr}
\makeatletter
\newcommand*{\addFileDependency}[1]{% argument=file name and extension
  \typeout{(#1)}
  \@addtofilelist{#1}
  \IfFileExists{#1}{}{\typeout{No file #1.}}
}
\makeatother

\newcommand*{\myexternaldocument}[1]{%
    \externaldocument{#1}%
    \addFileDependency{#1.tex}%
    \addFileDependency{#1.aux}%
}

\myexternaldocument{SI}

\begin{document}

\title{Elucidating the local atomic and electronic structure of amorphous oxidized superconducting niobium films}

\author{Thomas F. Harrelson}
\affiliation{Materials Science Division, Lawrence Berkeley National Laboratory, Berkeley, CA 94720, USA}
\affiliation{Molecular Foundry, Lawrence Berkeley National Laboratory, Berkeley, CA 94720, USA.}

\author{Evan Sheridan}
\affiliation{Materials Science Division, Lawrence Berkeley National Laboratory, Berkeley, CA 94720, USA}
\affiliation{Molecular Foundry, Lawrence Berkeley National Laboratory, Berkeley, CA 94720, USA}
\affiliation{ Theory and Simulation of Condensed Matter, Department of Physics, King's College London, The Strand, London WC2R 2LS, UK. }

\author{Ellis Kennedy}
\affiliation{Department of Materials Science and Engineering, University of California, Berkeley, CA 94720, USA}

\author{John Vinson}
\affiliation{Material Measurement Laboratory, National Institute of Standards and Technology, Gaithersburg, MD 20899, USA}
% New, poorly implemented NIST review requirement

\author{Alpha T. N'Diaye}
\affiliation{Advanced Light Source, Lawrence Berkeley National Laboratory, Berkeley, CA 94720, USA}

\author{M. Virginia P. Alto\'{e}}
\affiliation{Molecular Foundry, Lawrence Berkeley National Laboratory, Berkeley, CA 94720, USA}

\author{Adam Schwartzberg}
\affiliation{Molecular Foundry, Lawrence Berkeley National Laboratory, Berkeley, CA 94720, USA}

\author{Irfan Siddiqi}
\affiliation{Materials Science Division, Lawrence Berkeley National Laboratory, Berkeley, CA 94720, USA}
\affiliation{Department of Physics, University of California, Berkeley, CA 94720, USA}

\author{D. Frank Ogletree}
\affiliation{Molecular Foundry, Lawrence Berkeley National Laboratory, Berkeley, CA 94720, USA}

\author{Mary C. Scott}
\affiliation{Department of Materials Science and Engineering, University of California, Berkeley, CA 94720, USA}
\affiliation{NCEM, Molecular Foundry, Lawrence Berkeley National Laboratory, Berkeley, CA 94720, USA}

\author{Sin\'{e}ad M. Griffin}
\affiliation{Materials Science Division, Lawrence Berkeley National Laboratory, Berkeley, CA 94720, USA}
\affiliation{Molecular Foundry, Lawrence Berkeley National Laboratory, Berkeley, CA 94720, USA}

\begin{abstract}
Qubits made from superconducting materials are a mature platform for quantum information science application such as quantum computing. However, materials-based losses are now a limiting factor in reaching the coherence times needed for applications. In particular, knowledge of the atomistic structure and properties of the circuit materials is needed to identify, understand, and mitigate materials-based decoherence channels. In this work we characterize the atomic structure of the native oxide film formed on Nb resonators by comparing fluctuation electron microscopy experiments to density functional theory calculations, finding that an amorphous layer consistent with an Nb$_2$O$_5$ stoichiometry. Comparing X-ray absorption measurements at the Oxygen K edge with first-principles calculations, we find evidence of {\it d}-type magnetic impurities in our sample, known to cause impedance in proximal superconductors. This work identifies the structural and chemical composition of the oxide layer grown on Nb superconductors, and shows that soft X-ray absorption can fingerprint magnetic impurities in these superconducting systems.
\end{abstract}

\maketitle

%\section{Introduction}

%%%%%% Introduction: P1 - Context %%%%%% 

Superconducting qubits are one of the leading solid-state platforms for QIS (quantum information science) applications, with reported coherence times reaching $\sim$100 microseconds\cite{Devoret2013-mn,Kjaergaard_et_al:2020}. Despite this, materials-based decoherence channels contribute significantly to microwave losses, and are now a central hurdle in device coherence and scaling\cite{McDermott2009-qi}. In particular, the inevitable inhomogeneities that are present from growth and fabrication, such as interfaces, defects, and structural disorder, each contribute to the decoherence in qubits amde from superconducting materials\cite{Oliver2013-eo,De_Leon2021-je}. 

Precise knowledge of the atomistic structural and chemical makeup of superconducting qubit materials is particularly necessary for understanding materials-dependent decoherence processes. Intrinsic noise sources in superconducting qubits are typically classified into two categories -- two-level system (TLS) noise, and non-TLS noise\cite{muller_arxiv19}. TLSs are fluctuating two-level states comprising local energy minima in the atomic structural potential which were originally proposed to describe the microstructure of amorphous materials\cite{Phillips1987-jc}. TLSs can couple to electric and magnetic fields, reducing a qubit's coherence time. Since the amorphous materials present on superconducting qubit surfaces consist of a variety of bonding environments, TLSs can host a range of barrier heights and tunnelling rates, and correspondingly a distribution of fluctuation frequencies even within a given material\cite{Burnett2016-qt}. Because of this, characterization of the local atomic arrangements is needed to build any predictive description of TLS-related decoherence. 
%JTV: Does any of this need citations?

Non-TLS noise intrinsic in QIS materials includes the presence of nonequilibrium quasiparticles (QP)\cite{Wilen2020-ri, Cardani2020-xi}  and of magnetic impurities\cite{Kharitonov2012-dy, Proslier2011-ro, Sheridan_et_al:2021}. While careful shielding can mitigate some of these effects, the decay and control of QPs is materials dependent, and can be crucially influenced by nanofabrication and materials' control\cite{Vepsalainen2020-jl,Wilen2020-ri,Martinis2020-pn}. Another key non-TLS loss mechanism is Cooper pair breaking induced by the presence of magnetic impurities\cite{Kharitonov2012-dy}, which can occur due to materials' defects, interfaces, and surfaces, and cause impedance losses in the superconductor\cite{Kharitonov2012-dy, Proslier2011-ro, Sheridan_et_al:2021}. Therefore, to understand the structure-coherence relationships associated with the materials' properties in superconducting qubits, knowledge of the local structural and chemical environment is needed, regardless of the origin (TLS, non-TLS) of the noise.

Superconducting qubits are typically comprised of Al/AlO$_x$/Al Josephson junctions with superconducting circuit elements commonly made from Al, Nb, Ta, and alloys containing these\cite{Oliver2013-eo, Place2021-te}. Of these, Nb has many advantages over other superconducting materials including low kinetic inductance resulting in reduced variability, and a higher superconducting gap making it less susceptible to QP poisoning\cite{Kaplan_et_al:1976}. Importantly, Nb forms a relatively clean surface, and is a mature material for the advanced processing and lithographic patterning that is required for contemporary qubit fabrication and for future scaling of highly coherent superconducting architectures. However, Nb readily forms surface oxides such as NbO, NbO$_2$ and Nb$_2$O$_5$, which introduce both TLS and non-TLS losses in the qubit\cite{Delheusy_et_al:2008, altoe_arxiv20}.  Previous work has looked at the use of ultrahigh vacuum packing to reduce surface contamination\cite{Mergenthaler_et_al:2021}, in addition to an understanding of the influence of both oxide surface removal\cite{altoe_arxiv20} and regrowth\cite{Verjauw_et_al:2021} on the performance of superconducting resonators.

Despite extensive research on the variety of loss channels and their mitigation through surface treatments and fabrication~\cite{Romanenko2017-gy, Romanenko2020-wn}, the precise microscopic origins of TLS and non-TLS losses in superconducting systems is not known. This is primarily due to the difficulty in accessing information about the local structural and chemical environments which critically control the presence of these losses. Since the native oxides formed on Nb are often amorphous,  conventional diffraction and computational techniques cannot be used for structural information. Theoretical treatments often either rely on having crystalline materials with periodic boundary conditions\cite{Heinrich_et_al:2018}, or propose phenomenological models without incorporating nanoscale structural information.  Instead, in this work, we combine Fluctuation Electron Microscopy (FEM), X-Ray Absorption Spectroscopy (XAS), and first-principles calculations to investigate the structural and chemical composition of amorphous oxides on superconducting Nb films. We classify the short- and mid-range structural properties of our oxides by comparing our \textit{ab initio} calculations with experiments, identifying the structural and chemical makeup of surface Nb oxides on superconducting resonators. 

%\section{Methods}

To characterize the mid-range atomic structure of the amorphous films we used FEM, a 4-D scanning transmission electron microscopy technique that is sensitive to medium-range atomic ordering in disordered materials~\cite{voyles_2002}. FEM experiments were performed using an FEI TitanX operated at an acceleration voltage of 200 kV. Additionally, XAS measurements of the O K-edge were performed at the bending magnet beamline 6.3.1 at the Advanced Light Source at Lawrence Berkeley National Laboratory. We consider three different Nb treatments: (1) unpatterned, oxidized Nb films without any treatments, (2) Nb film from a chip patterned with qubits, and (3) Nb film from a chip with resonators only (no Josephson junctions), which allows us to potentially observe changes in the Nb oxides with these different fabrication steps (Table~\ref{tab:xas_samples}). XAS was performed on all three samples.  FEM was performed on Sample~2 because it had the thickest oxide layer, which was required for improved signal in FEM analysis. Further details of the FEM and XAS measurements are given in the SI. 

Electronic and magnetic properties were calculated using density functional theory (DFT) as implemented in the Vienna Ab initio Simulation Package (VASP)~\cite{kresse93}. We used Nb$_2$O$_5$ amorphous structures that were generated previously with \textit{ab initio} molecular dynamics as detailed in Ref.\cite{Sheridan_et_al:2021}, which are available on Zenodo\cite{zenodo_structures}.
X-ray absorption calculations were carried out using the Bethe-Salpeter equation (BSE) formalism as implemented within the {\sc ocean} code\cite{ocean1,*ocean2,*ocean0}. The BSE calculations use a basis of electron orbitals from DFT calculated with {\sc Quantum ESPRESSO}, \cite{espresso2,*espresso0} 
with pseudopotentials from the PseudoDojo collection. \cite{pspdojo1,*pspdojo0,PhysRevB.88.085117,*oncvp} More details on the DFT and XAS calculations are given in the SI.

%\section{Results}

\begin{figure}
    \centering
    \includegraphics[width=1.0\linewidth]{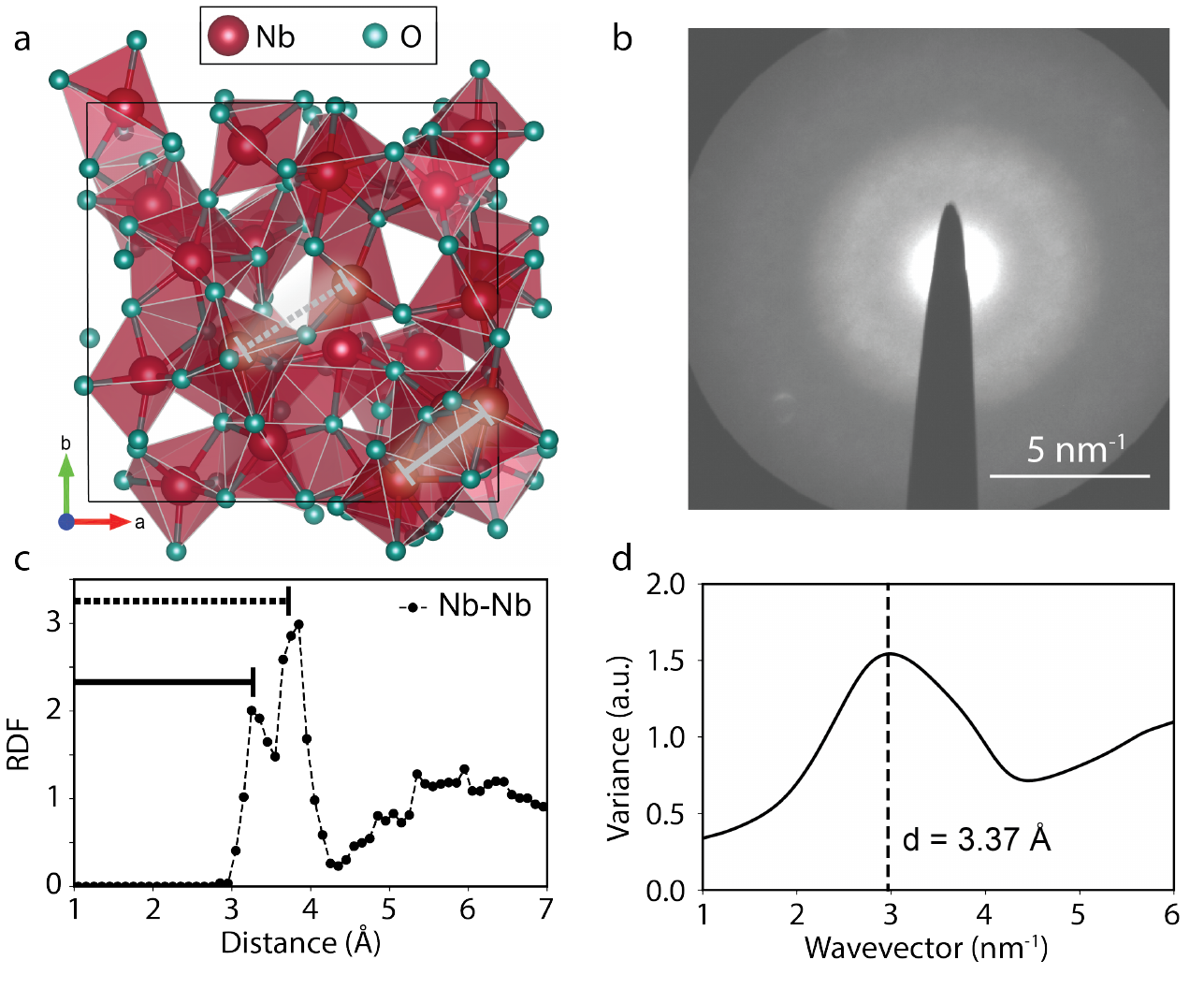}
    \caption{(a) Representative \textit{ab initio} molecular dynamics generated amorphous structure of Nb$_2$O$_5$. (b) Averaged speckle pattern  of Nb$_2$O$_5$ using FEM over many diffraction patterns. (c) Radial Distribution Function for Nb$_2$O$_5$ obtained from averaging over nine amorphous stoichiometric configurations generated using \textit{ab initio} molecular dynamics,  (d) Annular mean of normalized variances of the FEM data measuring the average interatomic spacing between Nb centers in the Nb$_2$O$_5$.}
    \label{fig:fig1}
\end{figure}

%\subsection{Local Atomic Structure from Fluctuation Electron Microscopy and Density Functional Theory}
We first describe our FEM diffraction results of a representative oxidized Nb sample with the largest oxide thickness (Sample 2), and compare the short-range structural description to \textit{ab initio} generated structures. In contrast to other diffraction techniques, which generally identify long-range ordering, FEM is uniquely sensitive to the medium-range ordering on the size scale of the electron beam probe \cite{Voyles_et_al:2003, daulton_2010}. FEM data is acquired by rastering a small electron probe over a sample and capturing a diffraction pattern at each probe location. The diffraction patterns are digitally preprocessed to remove imaging distortions, and the variance of the measured intensity as a function of scattering vector is calculated \cite{Kennedy_et_al:2020}. As Bragg scattering in the diffraction patterns creates large variance in intensity, the calculated variance is a metric for ordering in the amorphous material on the length scale of the electron probe \cite{hwang_2011}.  Full details of the FEM method and data analysis are given in the SI. In Figure~\ref{fig:fig1}b, we show the average speckle pattern of many  nanodiffraction patterns taken over the Nb oxide region of the film cross-section (see SI). The brighter spots in the speckled halo primarily represent Nb-Nb distances because electron scattering from Nb atoms dominates over scattering from O atoms. The broad  diffuse halo present in the average nanodifraction pattern suggests that the Nb oxide film is amorphous.  In  Figure~\ref{fig:fig1}(d) we show an average spatial variance computed from six regions of the sample, where each region differs in its thickness, as shown in Figure S1 of the Supplementary material. The broad peak centered at the wavevector $\approx$3~nm$^{-1}$ is a measure of the average interatomic spacing between Nb centres, corresponding to an average Nb-Nb distance of 3.37~{\AA}.

We next analyze {\it ab initio}-generated amorphous structures to investigate the short- and medium-range structural order across a sample of stoichiometric Nb$_2$O$_5$ amorphous configurations. Figure~\ref{fig:fig1}(a) illustrates a representative stoichiometric amorphous configuration of Nb$_2$O$_5$ containing 105 atoms in the unit cell. The solid line indicates the Nb-Nb distance for edge sharing Nb sites in Nb$_2$O$_5$, while the dashed line highlights the longer Nb-Nb  distances for corner sharing Nb sites. These features are also present in Figure~\ref{fig:fig1}(c), where we show the averaged radial distribution function (RDF) obtained from nine stoichiometric amorphous configurations of Nb$_2$O$_5$ whose volume and internal coordinates were optimized using DFT. We see from the first peak that the shorter edge sharing Nb sites are typically 3.15~{\AA} apart, and the longer corner sharing Nb sites are 3.8~{\AA} apart as indicated by the second peak. The immediate dip of the RDF at 4~{\AA} suggests that the edge- and corner-sharing environments shown in Figure~\ref{fig:fig1}(a) are the primary structural motifs present in our amorphous Nb$_2$O$_5$. Given the reasonable comparison between \textit{ab initio}-generated amorphous structures and FEM analysis of our Nb oxide thin films, we can conclude that indeed our films are amorphous, lacking any long-range order, and that our generated structures can be used for further analysis. Additionally, we find the average Nb-Nb distance in the FEM measurement to be 3.37 \AA, which is between the average corner- and edge-shared Nb-Nb distances in the \textit{ab initio} structures, suggesting our amorphous films comprise a mix of corner- and edge-sharing polyhedra.

%\begin{figure}
 %   \centering
 %   \includegraphics[width=1.0\linewidth]{Figures/fig2-FEMPCA.png}
 %   \caption{Blah blah}
  %  \label{fig2}
%\end{figure}

%\subsection{Electronic Structure from X-Ray Absorption Spectroscopy and Density Functional Theory}

\begin{figure}
    \centering
    \includegraphics[width=0.95\linewidth]{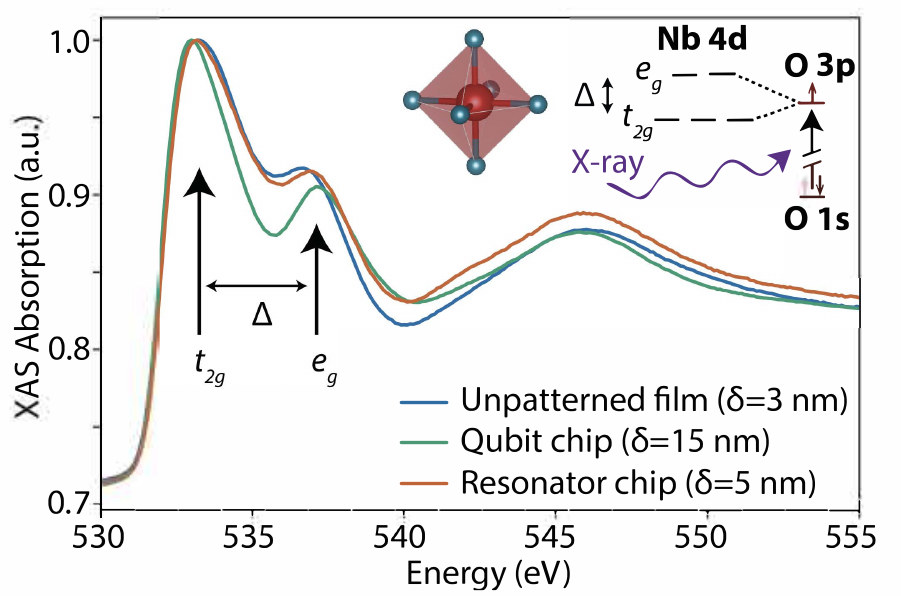}
    \caption{Measured XAS spectra of O K-edge for the three different samples described in Table~\ref{tab:xas_samples}. Inset: Sketch of the electronic structure of octahedrally coordinated Nb forming  $t_{2g}$ and $e_g$ split orbital sets. These hybridize with the unoccupied O orbitals that are excited upon X-ray absorption, creating the observed splitting between the peaks in the data. Spectra were normalized by matching the baselines, and dividing by the maximum value in the 525~eV to 550~eV window.}
    \label{fig3}
\end{figure}

We next use X-ray absorption spectroscopy (XAS) to obtain information about the local morphology, electronic structure, and potential magnetism. We focus on O K-edge spectra for three different NbO$_x$ samples, which are described in Table~\ref{tab:xas_samples}. 
While XAS of the O K edge  probes unoccupied {\it p}-type states surrounding the oxygen atoms, these states are hybridized with the neighboring Nb, and so provide information on both the Nb and O species. 
In Figure~\ref{fig3} we plot the measured XAS for the three samples, and find the XAS is similar for all three. As is typical in transition metal oxides, we identify the two peaks at 533~eV and 537~eV as hybridized with the empty Nb \textit{4d} orbitals, split by the crystal field splitting $\Delta$ into lower-energy $t_{2g}$ and higher-energy $e_g$ states. 
The broad feature near 544~eV reflects hybridization with Nb \textit{5sp}-like states. 
Changes in the relative intensities of the $t_{2g}$ and $e_g$, and splitting between them $\Delta$, and (less reliably) position of the edge onset, reflect changes in the Nb {\it d}-manifold occupation, strength of the Nb-O bonding, and oxidation state of the metal ion, respectively\cite{Frati_et_al:2020}.
%For the transition metal oxide structures considered here (NbO$_x$), we focus on the the O K-edge spectra. In Figure~\ref{fig3}, we show the O K-edge spectra for three different NbO$_x$ samples, which are described in Table~\ref{tab:xas_samples}. In this energy region, we find that there are two dominant peaks near 530~eV and 535~eV, along with a broader peak near 544~eV. 
%
%Importantly, probing the O K-edge spectra gives information about the corresponding Nb electronic structure. We see two dominant peaks in the XAS spectra corresponding to the average crystal field splitting of the $T_{2g}$ and $E_g$ \textit{d}-orbitals of an octahedrally-coordinated Nb atom. X-rays excite the O \textit{1s} orbital to the lowest unoccupied orbital projected on the O atom, which is closest in character to the O \textit{3p} orbital. This unoccupied orbital is hybridized with the $T_{2g}$ and $E_g$ \textit{d}-orbitals of the Nb atom which splits the O \textit{3p} orbital into two distinct hybridized orbitals, which are observed in the XAS spectrum (Figure~\ref{fig3}).  

%As a result, we expect that XAS spectra provide useful information about the local morphology and electronic structure, which can be correlated with changes in magnetic properties, particularly because it has recently been shown that local structural descriptors can be used to predict the presence/absence of magnetic impurities in amorphous NbO$_x$ structures~\cite{Sheridan_et_al:2021}.

\begin{table}%[h]
\centering
    \begin{tabular}{| M{0.13\linewidth} | M{0.17\linewidth} | M{0.6\linewidth} |} \hline
        Sample & NbO$_x$ Thickness & Description \\
        \hline \hline
        1 & 3~nm & Unpatterned, oxidized Nb film. \\ \hline
        2 & 15~nm & Nb film fabricated with qubits including AlO$_x$ Josephson junctions.\\ \hline
        3 & 5~nm & Nb film fabricated with resonators only (no Josephson junctions). \\\hline
    \end{tabular}
    \caption{Summary of sample details  used in  experiments. XAS measurements were performed on all three samples, whereas FEM measurements were performed on Sample 2.}
    \label{tab:xas_samples}
\end{table}

Comparing the XAS spectra of the three measured samples shows that the unpatterned film (Sample 1) and the resonator chip (Sample 3) are the most similar. We observe a slight increase in energy of the peak near 537~eV, and the slight increase in intensity of the broad feature near 544~eV for the resonator sample (Sample 3) compared to the unpatterned sample (Sample 1). The qubit sample (Sample 2) has the largest NbO$_x$ thickness ($\sim15$~nm), and largest increase in energy of the 537~eV peak. The observed increase in the energy of the 537~eV peak in the patterned samples suggests a greater crystal-field splitting hence more crystalline character compared to the unpatterned films. 

\begin{figure}
    \centering
    \includegraphics[width=0.95\linewidth]{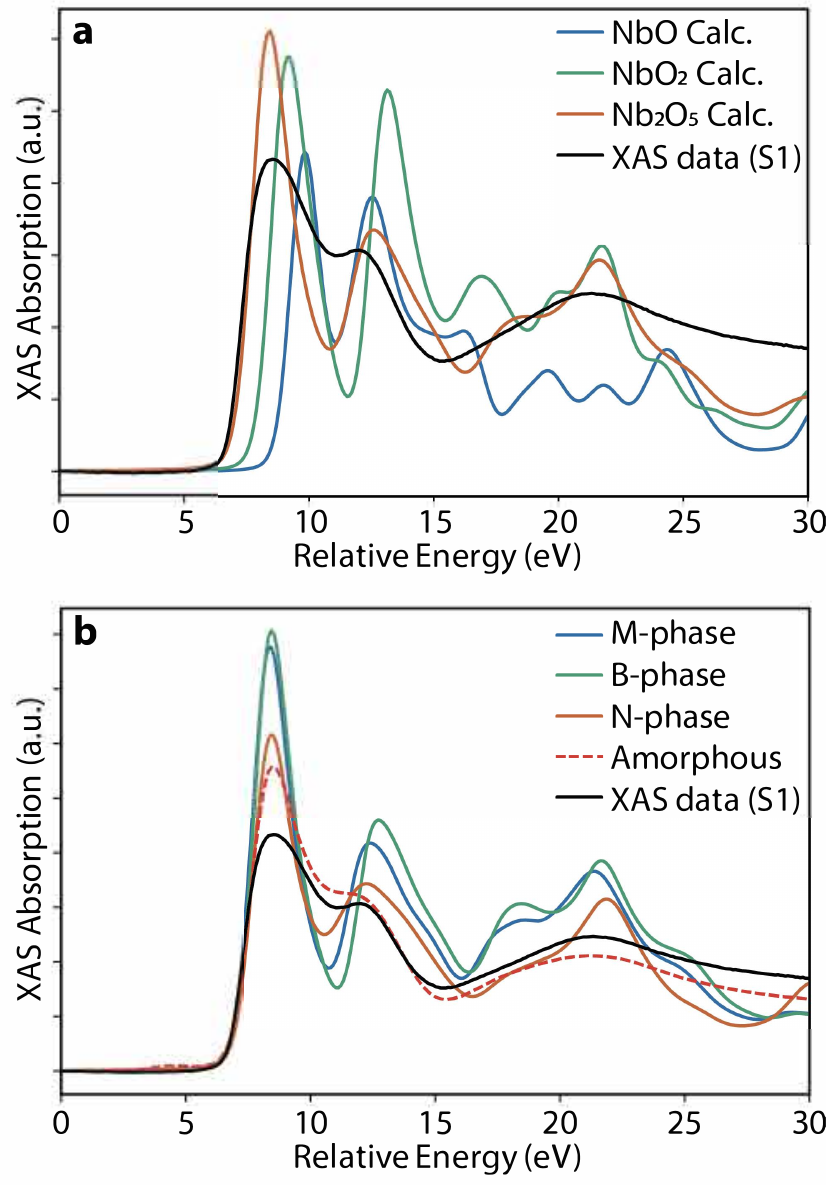}
    \caption{(a) Calculated XAS spectra for crystalline NbO, NbO$_2$ and Nb$_2$O$_5$ (averaged over all three calculated phases) and XAS measurements of the O K edge of Sample 1. (b) Calculated XAS spectra for crystalline Nb$_2$O$_5$ in the M-, B-, and B- phases, a representative \textit{ab initio} generated amorphous structure, and XAS measurements of Sample 1. Experimental data is normalized by rigidly shifting the spectrum to the relative scale, removing the background signal, and normalizing the heights to be comparable to our XAS calculations. }
    \label{fig4}
\end{figure}

We use a combination
of DFT and BSE calculations to further analyze the XAS spectra.  We calculate spectra for fifteen different Nb$_2$O$_5$ amorphous configurations (both stoichiometric and non-stoichiometric), and five different crystalline phases of Nb-oxides. In Figure~\ref{fig4}(a), we plot the calculated crystalline spectra for NbO ($Pm3m$), NbO$_2$ (\textit{$P4_2/mnm$}), and the average of 3 different Nb$_2$O$_5$ phases (N-phase ($C_2/m$), M-phase ($I4/mmm$), and B-phase ($C_2/c$) and compare to the experimental spectrum of Sample 1.  We find that the measured XAS spectra are best described by Nb$_2$O$_5$. The splitting between the two dominant peaks is larger in the crystalline reference samples, while the relative heights of the two dominant peaks is  qualitatively described by Nb$_2$O$_5$, suggesting amorphous structures with  Nb$_2$O$_5$ stoichiometry. We further find that as the oxidation state of the Nb atom increases ($+2$ in NbO to $+5$ in Nb$_2$O$_5$), both the intensity of the first peak increases, and the ratio of the intensity of the first peak to the second peak increases. 
%JTV1: Not all of the Nb in these simulations are octohedral (?) should we rephrase the following
This is partially explained by considering the resulting filling of the  $t_{2g}$ and $e_g$ states of an octahedrally coordinated Nb atom (see inset of Figure~\ref{fig3}); NbO deviates slightly from the trend because the coordination of the Nb atoms is square planar.

%In Figure~\ref{fig4}b, we compare the calculated XAS spectra for three different crystalline polymorphs and an amorphous structure of stoichiometric Nb$_2$O$_5$ to identify the most likely structure in our films. We find that both the average amorphous spectrum and the crystalline $N$-phase spectrum are most similar to the experimental spectrum from Sample 1.  We choose Sample 1 since we anticipate the oxidized film with no additional fabrication steps is most similar to a completely amorphous phase.  
In Figure~\ref{fig4}(b), we compare the calculated XAS spectra for three different crystalline polymorphs and an amorphous structure of stoichiometric Nb$_2$O$_5$ to the measured XAS of Sample 1. 
We choose Sample~1 since we anticipate the oxidized film with no additional fabrication steps is most similar to a completely amorphous phase.  
We find that both the average amorphous spectrum and the crystalline N-phase spectrum are most similar to the experimental spectrum from Sample 1. Of the crystalline phases, we find that the N-phase best agrees with the XAS measurements, but the calculation shows a larger splitting between the two dominant peaks than the measured spectrum. This is the case for all of the considered crystalline phases of Nb$_2$O$_5$ (Figure~\ref{fig4}(b)), which is caused by the crystalline order increasing the crystal field splitting.

%This is explained by the known underestimation of band localization using semi-local density functionals -- we remedy this using an energy renormalization factor of 1.144 by fitting our calculated amorphous spectra to Sample 1 XAS measurements.
%JTV need to adjust how we talk about the energy stretch. If all the shown figures are with the stretch, I think we should move the discussion of it to the SI

%\subsection{Magnetic Impurity States in XAS Spectra}

\begin{figure}
    \centering
    \includegraphics[width=0.95\linewidth]{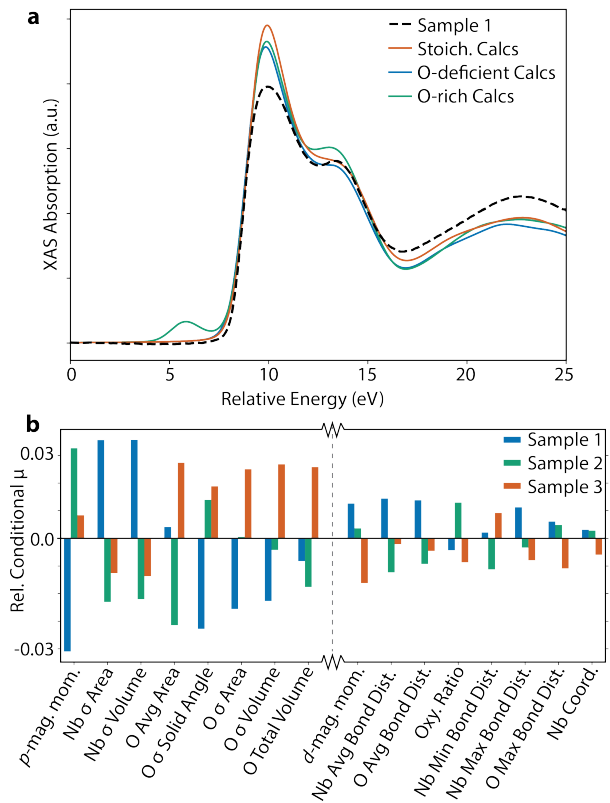}
    \caption{(a) Comparison between  stoichiometric, oxygen-deficient, and oxygen-rich amorphous calculated spectra versus the experimental spectrum of sample 1. (b) Statistical analysis of the expected relative changes in structural descriptors given the three experimental XAS spectra. The highest variance descriptors are on the left and lowest variance descriptors are on the right.}
%    \includegraphics[width=0.8\linewidth]{Figures/xas_mag-01.png}
%    \caption{(a) Expectation of \textit{p}- and \textit{d}-type magnetic impurity XAS spectra. (b) Average ($\mu$) and standard deviation ($\sigma$) of all calculated XAS spectra is compared with the three experimental spectra. The vertical dashed lines represent the region in which the calculated spectra can reasonably explain the experimental data.}
    \label{oxygen}
\end{figure}

Previous works suggest magnetic impurities contribute to impedance-based losses in superconducting qubits\cite{Kharitonov2012-dy, Sheridan_et_al:2021}. In particular, $d$-type magnetic impurities on  Nb atoms were found to be more detrimental than $p$-type impurities on O atoms in Nb oxides\cite{Sheridan_et_al:2021}. To investigate if our XAS measurements can identify a low density of magnetic impurities, we compare our calculated XAS spectra with those measured. We divide our calculations into three groups; stoichiometric Nb$_2$O$_5$, oxygen rich Nb$_2$O$_5$ (includes oxygen interstitials or Nb vacancies), and oxygen poor Nb$_2$O$_5$ (includes oxygen vacancies) with the results given in Figure \ref{oxygen}.  As expected, we find a pre-edge feature in the oxygen rich calculations coming from O-O dangling bonds, and resulting in the \textit{p}-type magnetic impurities. However, such a pre-peak feature is not observed in any of the measured XAS spectra, so we can conclude that there is not a significant density of \textit{p}-type magnetic impurities in our measured samples. We find a slightly better agreement between the oxygen-poor calculated spectra and the measured spectra with shape of the second peak at $\approx 13$~eV a closer match in this case. This suggests the presence of \textit{d}-type magnetic impurities associated with oxygen poor (and Nb rich) samples. 
To further quantify this, we perform statistical analysis on the 1250 distinct calculated atomic spectra and compare them to the measured XAS to correlate spectral changes with structural and magnetic changes (details in the SI). These results suggest that small densities of \textit{d}-type magnetic impurities are present in all three measured samples with an estimated density of 1.8$\times10^{22}$, 1.7$\times10^{22}$, and 1.5$\times10^{22}$ \textit{d}-type impurities per mol formed from oxygen vacancies found in Samples 1, 2, and 3, respectively. 
%JTV T and TT come out of nowhere and might confuse a reader
This density of magnetic moments is  higher than previous magnetic measurements on bulk crystalline \textit{T}- and \textit{TT}-Nb$_2$O$_5$ where they estimated a density of $10^{21}$ -- $10^{22}$ effective magnetic moments per mol\cite{herval_15}. However, the greater density of magnetic moments in our samples is consistent with the increased disorder and off-stoichiometry that we expect in our amorphous surface oxides. To elucidate the structural changes amongst the different samples that are correlated with changes in the XAS spectra, we calculate the conditional mean for each structural and magnetic descriptor given to the experimental spectrum. The relative expected changes for the most and least varying descriptors are plotted in Figure~\ref{oxygen}b. The shape descriptors (volume, area, etc.) refer to Voronoi polyhedra constructed around each atom, $\sigma$ values refer to distortions within those descriptors (full descriptions are in the SI). We find that the fabrication procedure has a reasonably large effect on the \textit{p}-type impurity density along with the shape descriptors of both Nb and O atoms. Bond length and coordination characteristics along with \textit{d}-type impurity density showed less variation amongst the samples. 
In summary,our FEM measurements confirm the lack of long-range order in a representative Nb-oxide film, observed from the broad halo in the average speckle pattern of the FEM image. Comparing the calculated RDF of our \textit{ab initio}-generated stoichiometric Nb$_2$O$_5$ amorphous structures to the angular average of the FEM pattern indicates that our generated amorphous configurations are a good representation of the distribution of structures observed in real Nb-oxide films, containing a mix of edge and corner-sharing polyhedra motifs. We next compared our measured XAS spectra for a selection of Nb-oxide samples (Table \ref{tab:xas_samples}) to first-principles calculations of both crystalline and amorphous Nb-oxide compounds, of which the amorphous phase most closely matched the data, which isconsistent with our FEM results, and prior elemental analysis of Nb-oxide films\cite{altoe_arxiv20}. Finally, we analyze our first-principles predictions for signatures of magnetic impurities in the amorphous configurations to identify experimental markers of these magnetic impurities in the XAS spectra. We find a better fit of the XAS spectra for  Nb$_2$O$_5$ configurations with oxygen vacancies, suggesting the presence of \textit{d}-type magnetic impurities. We find no evidence for pre-edge impurity states associated with \textit{p}-type magnetic impurities. Our results give an estimate of the density of decoherence-inducing local magnetic moments, and suggest experimental fingerprints for  the characterization of superconducting thin films using spectroscopic approaches. 

%We find that the average Nb-Nb distance in the FEM measurements is 3.37 \AA, lying between the average values of the calculated average Nb-Nb separation in the \textit{ab initio} structures of 3.15 \AA\ (edge sharing), and 3.8 \AA\ (corner sharing). From this we conclude that the measured amorphous thin film has a combination of both edge- and corner-sharing polyhedra, consistent with  \textit{ab initio} calculations. 

%We next compared our measured XAS spectra for a selection of Nb-oxide samples (Table \ref{tab:xas_samples}) to first-principles calculations of both crystalline and amorphous Nb oxide compounds. We find the best agreement to all three of the measured samples to be for the amorphous Nb$_2$O$_5$ phase, consistent with our FEM measurements and previously reported elemental analysis of Nb oxide thin films\cite{altoe_arxiv20}. 

%Finally, we analyze our first-principles predictions for signatures of magnetic impurities in the amorphous configurations to identify experimental markers of these magnetic impurities in the XAS spectra. We find a better fit of the XAS spectra for  Nb$_2$O$_5$ configurations with oxygen vacancies, suggesting the presence of \textit{d}-type magnetic impurities. We find no evidence for pre-edge impurity states associated with \textit{p}-type magnetic impurities. Our results give an estimate of the density of decoherence-inducing local magnetic moments, and suggest experimental fingerprints for  the characterization of superconducting thin films using spectroscopic approaches. 

\section*{Data Availability}
The data that support the findings of this study are openly available in Zenodo at Ref~\cite{zenodo_xas}.

\section*{Acknowledgments}
We thank John Clarke and David Santiago for useful discussions.  Specific software and hardware is identified for information purposes only and is not intended to imply recommendation or endorsement by NIST. 
This work was funded by the U.S. Department of Energy, Office of Science, Office of Basic Energy Sciences, Materials Sciences and Engineering Division under Contract No. DE-AC02-05-CH11231 ``High-Coherence Multilayer Superconducting Structures for Large Scale Qubit Integration and Photonic Transduction program (QIS-LBNL)". 
This research used resources of the National Energy Research Scientific Computing Center (NERSC), a U.S. Department of Energy Office of Science User Facility located at Lawrence Berkeley National Laboratory, operated under Contract No. DE-AC02-05CH11231. E.S. acknowledges support from the US-Irish Fulbright Commission, the Air Force Office of Scientific Research under award number FA9550-18-1-0480 and the EPSRC Centre for Doctoral Training in Cross-Disciplinary Approaches to Non-Equilibrium Systems (EP/L015854/1). This work also used the Extreme Science and Engineering Discovery Environment (XSEDE), which is supported by National Science Foundation grant number ACI-1548562. Electron microscopy data acquisition for this work was supported by National Science Foundation STROBE Grant No. DMR-1548924. Work at the Molecular Foundry was supported by the Office of Science, Office of Basic Energy Sciences, of the U.S. Department of Energy under Contract No. DE-AC02-05CH11231. This research used resources of the Advanced Light Source, which is a DOE Office of Science User Facility under Contract no. DE-AC02-05CH11231. 

\section*{Author Declarations}
The authors have no conflicts to disclose.

\bibliography{refs}

\end{document}